# In silico clinical trials in drug development: a systematic review

Bohua Chen[1]; Lucia Chantal Schneider[1]; Christian Röver[1];
Emmanuelle Comets[2]; Markus Christian Elze[3]; Andrew Hooker[4];
Joanna IntHout[5]; Anne-Sophie Jannot[6]; Daria Julkowska[7]; Yanis Mimouni[8];
Marina Savelieva[9]; Nigel Stallard[11]; Moreno Ursino[13];
Marc Vandemeulebroecke[5,10]; Sebastian Weber[9]; Martin Posch[12]; Sarah Zohar[13];
Tim Friede[1,14,15,16]

[1] Department of Medical Statistics, University Medical Center Göttingen, Göttingen, Germany.

[2] Irset (Institut de Recherche en Santé, Environnement et Travail) - UMR_S1085, Univ Rennes, Inserm, EHESP, Rennes, France.

[3] Data Sciences, F. Hoffmann-La Roche AG, Basel, Switzerland.

[4] Department of Pharmacy, Uppsala University, SE-751 24, Uppsala, Sweden.

[5] Department of IQ Health, Radboud university medical center, P.O. Box 9101, 6500 HB Nijmegen, The Netherlands.

[6] Université Paris Cité, HeKA, INRIA Paris, Inserm, Centre de Recherche des Cordeliers-Université Paris Cité, Paris, France.

[7] European Joint Programme on Rare Diseases, Institut Thématique Génétique, Génomique et Bioinformatique, INSERM, 75013 Paris, France.

[8] Thematic Institute of Genetics, Genomics & Bioinformatics, INSERM, Paris 75013, France.

[9] Novartis Pharma AG, Basel, Switzerland.

[10] Clinical Statistics & Analytics, Bayer Pharma AG, Basel, Switzerland

[11] Warwick Clinical Trials Unit, Warwick Medical School, University of Warwick, Coventry, CV47AL, UK.

[12] Medical University of Vienna, Center for Medical Data Science, Institute of Medical Statistics, Vienna, Austria

[13] Inserm, U1346, Université Paris Cité, Centre Inria Paris, HeKA, Paris, France

[14] DZHK (German Center for Cardiovascular Research), partner site Lower Saxony, Göttingen Germany

[15] DZKJ (German Center for Child and Adolescent Health), Göttingen, Germany

[16.] CAIMed – Lower Saxony Center for Artificial Intelligence, Göttingen, Germany

Corresponding author: Bohua Chen, bohua.chen@med.uni-goettingen.de

Humboldtallee 32, 37073 Göttingen, Germany

ABSTRACT

In the context of clinical research, computational models have received increasing attention over the past decades. In this systematic review, we aimed to provide an overview of the role of so-called *in silico clinical trials* (ISCTs) in medical applications. Exemplary for the broad field of clinical medicine, we focused on *in silico* (IS) methods applied in drug development, sometimes also referred to as *model informed drug development* (MIDD).

We searched PubMed and ClinicalTrials.gov for published articles and registered clinical trials related to ISCTs. We identified 202 articles and 48 trials, and of these, 76 articles and 19 trials were directly linked to drug development. We extracted information from all 202 articles and 48 clinical trials and conducted a more detailed review of the methods used in the 76 articles that are connected to drug development. Regarding application, most articles and trials focused on cancer and imaging-related research while rare and pediatric diseases were only addressed in 14 articles and 5 trials, respectively. While some models were informed combining mechanistic knowledge with clinical or preclinical (in-vivo or in-vitro) data, the majority of models were fully data-driven, illustrating that clinical data is a crucial part in the process of generating synthetic data in ISCTs. Regarding reproducibility, a more detailed analysis revealed that only 24% (18 out of 76) of the articles provided an open-source implementation of the applied models, and in only 20% of the articles the generated synthetic data were publicly available.

Despite the widely raised interest, we also found that it is still uncommon for ISCTs to be part of a registered clinical trial and their application is restricted to specific diseases leaving potential benefits of ISCTs not fully exploited.

Key words: In silico, systematic review, clinical trial, pediatric disease, rare disease, Model-Informed Drug Development

## INTRODUCTION

The development of new drugs is both costly and time-consuming [1]. Among other measures, the implementation of sophisticated and appropriate quantitative methodology in the development process may enhance the efficient use of resources. Computational models have evolved from a mere alternative data source (e.g. to clinical trials) to providing a toolbox for the efficient development of new medicines, for drug maintenance on the market, and the extension of indications for existing drugs (i.e. their re-purposing in completely new indications) [2].

*In silico clinical trials* (ISCTs) prove particularly useful in studies in contexts where researchers struggle to enroll sufficient numbers of participants, like studies addressing rare or pediatric diseases. The generated synthetic data may augment data from clinical studies, thus enhancing the evidence base. The availability and appropriate use of methods to synthesize different types of evidence then becomes key to an impartial and accurate assessment of the overall evidence [3]. In particular, different types of evidence, as derived from randomized experiments, real-world data, or computational models needs to be treated with appropriate consideration of their exact relationships and potential biases to allow for robust and reliable conclusions [4]. Regulatory agencies such as U.S. Food and Drug Administration (FDA) and European Medicines Agency (EMA) have already considered issues in extrapolation methods for pediatric populations by issuing dedicated guidelines [6, 7].

A variety of model-based approaches and applications have recently been defined as Model-Informed Drug Development (MIDD) by regulatory agencies and industry [5]. In particular, ISCTs can be used to supplement or substitute real patients with model-based simulations as part of MIDD. MIDD, as defined in the ICH-MIDD road map [5] and considered in the present review, involves more than just pharmacokinetic/pharmacodynamic (PK/PD) models.

The present article aims to review the impact of computational models in clinical medicine over the past decades, with a particular focus on drug development. To this end, we performed a systematic review aiming at journal publications describing or applying *in silico* (IS) methodology, and at registered clinical trials related to ISCTs. We sought to provide a comprehensive overview of the recent development of ISCTs, while also exploring differences in definitions used by various institutions. We outline the application of ISCTs across various diseases, especially rare and pediatric diseases, and provide detailed examples of the use of computational models in these specific contexts. In addition, we also consider methodological aspects, including the different types of computational models, the data sources used to inform them, and the reproducibility of published results.

The remainder of this article is structured as follows. In the following “Methods” section, the terminology and methods used in the systematic review are described, while the characteristics of the publications and trials identified are summarized in Section “Results”, including a closer look at three exemplary cases from the review and how IS methods were utilized. Finally, the paper closes with a discussion.

# METHODS

## Terminology

The exact definition of an ISCT varies between fields of application and between regulatory agencies. For example, the International Organization for Standardization (ISO) characterized an ISCT by “[...] the use of individualized computer simulation in the development or regulatory evaluation of a medicinal product, medical device or medical intervention” [8]. A similar, more detailed definition was given by the U.S. Food and Drug Administration (FDA) [9]; they defined an ISCT as “[...] an emerging application of computer modeling and simulation (CM&S) where device safety and/or effectiveness is evaluated using a ‘virtual cohort’ of simulated patients with anatomical and physiological variability representing the indicated patient population”. The purpose of ISCTs is described to include “[...] augmenting or reducing the size of a real-world clinical trial, providing improved inclusion-exclusion criteria, or investigating a device safety concern for which a real-world clinical trial would be unethical” [9]. Most recently, in the white paper by Musuamba et al. [2], ISCTs are defined as a “class of trials for pharmacological therapies or medical devices based on modeling and simulation technologies”. Furthermore, it is stated that the ISCTs’ purpose is to “[...] produce digital evidence that can serve in complement to or replacement of in vivo clinical trials for the development and regulatory evaluation of medical therapies” [10].

The common element in all of these definitions is the use of computer modeling and simulations to evaluate a diagnostic device or therapy. In contrast to the ISO, who kept their definition quite broad, Musuamba et al. and the FDA specifically characterized the purpose of an ISCT as to complement evidence from clinical trials. In the following, we adhere to the brief and concise definition of ISCTs given by Musuamba in 2021 [2]. Moreover, we clearly distinguish between data analysis, which plays a distinct role in modern clinical medicine, and data generation, which typically involves simulating biological systems and patients' outcomes rather than observing them experimentally.

The set of rules or the algorithm describing the behavior of the biological system is called an IS model if it is implemented and studied computationally [2]. IS models play a crucial role in generating digital evidence and since we later focus on drug development as one exemplary application field of ISCTs, we briefly outline some of the terminology and models frequently used in the field of model-informed drug development. The term IS model refers to a broader class of computational models ranging from fully mechanistic to fully data-driven models and encompasses deterministic models as well as stochastic models. Mechanistic models are typically deterministic and defined by a set of theoretical rules and algorithms based on known or hypothesized mechanisms, while data-driven models are developed from observations or data with the aim of inferring a set of rules explaining those data [2]. For example, pharmacokinetic (PK), pharmacodynamic (PD) and PK/PD models may describe the dose-concentration, the concentration-response and the dose-response relationship of a given drug through ordinary differential equations and rely to a varying degree on data to estimate parameters or even solely use those known from literature. A more comprehensive study of PK, PD and PK/PD models can be found in [11,12]. Extensions of the classic PK model, called physiologically-based pharmacokinetics (PBPK) and physiologically-based biopharmaceutics modeling (PBBM), are extensively explained in [13] and [14], respectively. A crucial aspect of successful model development is a solid understanding the underlying disease. Disease progression models describe how the disease dynamics are modified under the drug intake and can therefore be used to complement a PK/PD model. A good description of disease progression models and their use is provided in [15]. Quantitative systems pharmacology (QSP) is broadly defined as an approach to translational medicine that integrates PK, PD and disease progression modeling and aims at elucidating and validating new pharmacological concepts as well as applying them to the drug development process [2]. A review of QSP modeling cases can be found in [16]. To move from a patient-centered perspective to a population based approach and allow some variability in the effect of a drug, the corresponding PK and PD parameters may be modeled as random variables. This class of models is called population-based PK (popPK) or population-based PK/PD (popPK/PD), depending on which relationship is considered, and always requires patient data in addition to the mechanistic knowledge behind the PK and PD mechanisms to infer the underlying probability distribution. A more detailed explanation is given by [17]. Models based on artificial intelligence, e.g., machine learning or deep learning, heavily depend on the training data and do not necessarily reflect the actual underlying mechanisms accurately so that extra caution is necessary when extrapolating beyond the training data[2]. It is important to note that there is no generally accepted framework on how to differentiate all these models and their scope of application clearly and without overlap. An overview over some of the above mentioned model terms can be found in [2,9].

**Systematic literature search and trial selection**

We restricted our search to one literature database and one clinical trials registry. For both we chose the most popular ones (i.e. PubMed and clinicaltrials.gov, respectively), in order to review the methodology of ISCTs and their impact on clinical medicine.

<u>Search and screen process for journal articles (via PubMed)</u>

Starting with PubMed, we searched for the phrases *"in silico clinical trial"* or *"virtual clinical trial"* in articles published up to December 31, 2023. During the initial screening process of the resulting abstracts, we evaluated the relevance of the retrieved papers to our systematic review using the definition of ISCTs by Musuamba et al. 2021 [2]. At least two reviewers independently reviewed the abstracts and any disagreement was resolved by a third reviewer. We specifically excluded papers where computers were only used for image processing, endpoint evaluation, or treatment decision support, as these do not aim to reduce patient exposure to medical devices or therapies. The term "virtual trial" was sometimes used to refer to study designs where patients participate remotely, so papers referring to that meaning of the term *"virtual"* were also excluded. We further restricted our analysis to articles with a strong connection to ISCTs, eliminating articles that prepare for future ISCTs or simply validate the results of ISCTs. In a second step, we classified the remaining papers into three categories: *"review"*, *"application"* and *"methods"* in order to understand the field of research better. A *"review"* article gives a comprehensive summary of a broad or specific topic, an "*application*" article reports the protocol or results of actual ISCTs, and a "*methods*" article focuses on the IS model itself. The full list of included articles is provided in the supplementary material online, in *Supplementary material S1*.

As we are particularly interested in drug development and the methodology used in this application field of ISCTs we screened the remaining articles regarding their connection to drug development. We analyzed the methodological aspects like the used model and the reproducibility of results by considering only articles with a clear connection to a specific drug compound. This included the development of novel drugs as well as trials related to dose finding, drug re-purposing and new combinations of existing drugs. Papers without a link to these topics or those that did not mention a specific drug compound are excluded. Typical examples of excluded articles are related to imaging techniques and surgical methods as well as to the development of models that can be used for drug development in the future without naming any specific drug.

Search and screen process for registered clinical trials (via ClinicalTrials.gov)

We searched the ClinicalTrials.gov database for registered clinical trials beginning before December 31, 2023 using the search phrase *"in silico"*. At least two authors independently reviewed the study description and a third reviewer was consulted in cases of disagreement. In general, we included trials that applied IS methodology in all areas of medicine, regardless of study design or study population. Trials that do not perform an ISCT itself, but aim to lay the groundwork for future simulations by collecting data to inform IS models or evaluate results from previous ISCTs fall into the category "Preparation for future ISCTs" and "Validation of ISCT results", respectively. Trials in one of these categories are not excluded but considered relevant for further analysis and therefore the inclusion and exclusion criteria regarding the relevance of trials differ from those applied to the PubMed articles. Trials that incorporate an ISCT at some point in their process are classified as "application". The categories "methods" and "review" were not applied in the context of clinical trials. The full list of included clinical trials is provided in the supplementary material online, in *Supplementary material S2*.

**Data extraction and analysis**

Analysis of journal articles

We extracted general and application-related information for all journal articles classified as relevant no matter their relation to the field of drug development and considered only articles with a direct connection to drug development, i.e., articles that target a specific drug compound, for methodology-related information.

- Country and publication year

We used the publication date to investigate the interest in ISCTs over the past 30 years and the country of the first author's affiliation to analyze geographical aspects. To take into account the increase in the number of publications in general, we calculated the proportion of IS related articles among all articles with the keyword *"clinical trials"* for the corresponding year (see Figure 1).

● Disease classification

To determine which diseases IS modeling was applied to, we used the International Classification of Diseases, 11th Revision (ICD-11), which is divided into 26 chapters, each representing a distinct disease category. Most chapter names are self-explanatory, e.g., chapter 2 *"Neoplasms"* contains all diseases characterized by uncontrolled and abnormal growth of cells. Chapters that need further clarification are *"Factors influencing health status or contact with health services"* and *"Extension codes"*. The first one describes screenings and preventive examinations like CT scans and X-rays e.g. mammographies, the latter provides additional context to diagnoses, such as indicating that a condition is related to a medication, e.g. adverse reactions. We also added the category "General diseases" in the case the article did not mention a specified disease e.g. the article only describes the modeling of the immune system in general and not in response to a specific infection.

● Rare or pediatric disease context

As ISCTs can be used to reduce the number of trial participants, they promise to be particularly useful in the contexts of rare and pediatric diseases. In this review we refer to a disease as rare, if it affects less than 1 in 2000 people, which is the European Commission's definition [18], or when it has an entry in the Orpha.net database. We refer to a disease as pediatric, when it is only or mostly prevalent in people younger than 18 years.

● Relation to drug development

Further, we wanted to gain a deeper understanding of which particular IS modeling approaches were used in the field of drug development. We categorized the models as reported by the authors if the model class was specifically stated. In cases where the model was only referred to as mathematical or computational, we classified it according to the terminology section, if possible. The resulting 7 categories are: PK/PD, PBPK, PBBM, disease progression, QSP, popPK and other models. The latter category contained all the articles that could not be classified by us due to missing information on the model. In a second step, we assessed whether artificial intelligence (AI) techniques were integrated into any of the pharmacological models. To this end, we evaluated the modelling approaches against the AI definitions provided in Friedrich et al. (2022 [30], 2021 [31]).

● Data source for calibration and validation

All ISCTs rely on some kind of previously gathered data or knowledge to inform the models. Therefore, we also categorized the data sources used for calibration and validation if this information was available. Some commercial models or open-source implementations were ready to use and thus did not need calibration or validation. Consequently, we do not focus on the absolute number but rather the relative proportion of the different data sources. For the source of information used in the process of model calibration we distinguished between mechanistic knowledge, pre-clinical, clinical and registry data. For example a fully data driven model would be calibrated using only pre-clinical and/or clinical data in contrast to a mechanistic model that is mainly based on knowledge about the biological mechanisms and uses data only on some degree or even not at all. In contrast, the model validation always requires some form of data which we classify as either pre-clinical, clinical, registry or simulation data from other models outcomes (e.g. virtual cohorts).

- Model implementation and data availability

Finally, we consider the reproducibility of the published results, where we differentiate between implementation and data availability. If the implementation is available, we further sub-categorise the articles in *"commercial"* if a commercial platform or program like Simcyp™ PBPK Simulator is used, or *"open-source"* if the code is openly accessible. When no implementation is available, we distinguish between *"model fully specified"* and *"not available"*, where articles that at least contain the model equations and necessary parameter values are separated from the ones that did not even publish the full model specifications. For data availability, we focused on whether the data simulated in the ISCT and leading to the results stated by the authors, is openly available, available only on request, or not available.

<u>Analysis of registered clinical trials</u>

- Study start, country

For all clinical trials considered relevant, we extracted general and disease related information. The recorded start of the trial or in the case of future trials the estimated study start is used to analyze the increase in the number of ISCTs conducted in the past 30 years. In order to adjust for the increasing number of registered trials in general, we calculated the proportion of IS related trials among all registered clinical trials in the corresponding year (Figure 1). Furthermore, we considered the countries where the trials are carried out. For multi-center trials all countries were included.

- ICD-11 disease classification, rare and pediatric diseases

We used the International Classification of Diseases, 11th Revision (ICD-11) to determine in which disease areas IS methods are being applied, also focusing on whether targeted diseases are rare or pediatric. The categories and the corresponding in- and exclusion criteria used here are the same as in the case of the journal articles except that the category of "General diseases" did not apply because clinical trials generally target a specific disease.

- Relation to drug development

We classified the trials regarding their relation to drug development. In contrast to the journal articles, it suffices to distinguish between "no connection" and a specific drug (i.e., a drug compound). Since the trial descriptions generally do not elaborate on the methods used to analyze the data, we omit the categorization of models and data sources for calibration and validation as well as the evaluation of the reproducibility.

- Overlap information (NCT number in PubMed and PMID number in trials)

After analysing the data from the PubMed and ClinicalTrials.gov searches separately, we evaluate the overlap between these two data sources to address the questions if ISCTs described in a PubMed article are also registered in ClinicalTrials.gov and what kind of trials are cited in the paper. To that end, we searched every included journal article from the PubMed search for the phrase "NCT", as every clinical trial registered in ClinicalTrials.gov has a unique NCT identifier. Inversely, we also searched every included clinical trial from ClinicalTrials.gov for a PMID number to find articles cited in the study description.

Further details on the categories used in the classification described above and their inclusion and exclusion criteria can be found in *Categories_Supplementary*. Study characteristics are presented as descriptive summaries and R version 4.1.1 (R Foundation for Statistical Computing, Vienna, Austria) was used for all statistical analyzes and plots.

## RESULTS

### Included articles and trials

According to our chosen definition of ISCTs, we identified 202 articles and 48 trials for the final analysis. The full lists of included articles and registered trials are provided in the *Supplementary material S1* and *Supplementary material S2*. The top contributing journals from PubMed are *Journal of Medical Imaging (Bellingham)* and *Medical Physics*, contributing 17 articles (5.6%) each, followed by *Physics in Medicine and Biology* with 9 articles (3.0%), while moderate contributions come from *Computer Methods and Programs in Biomedicine* and *Scientific Reports* with 8 articles (2.6%) and *Annual International Conference of the IEEE*

*Engineering in Medicine and Biology Society* and *Journal of Diabetes Science and Technology* with 7 articles (2.3%). For more information regarding the journals please refer to Figure S1 in *Figures_Supplementary*. Among the trials identified through the search of ClinicalTrials.gov, only 1 (2%) had results, and 14 (29.2%) were completed. Further details can be found in Figures S2 and S3 in *Figures_Supplementary*. The search and screening process is summarized according to the PRISMA statement [19] and depicted in the PRISMA flow charts (Figure 1).

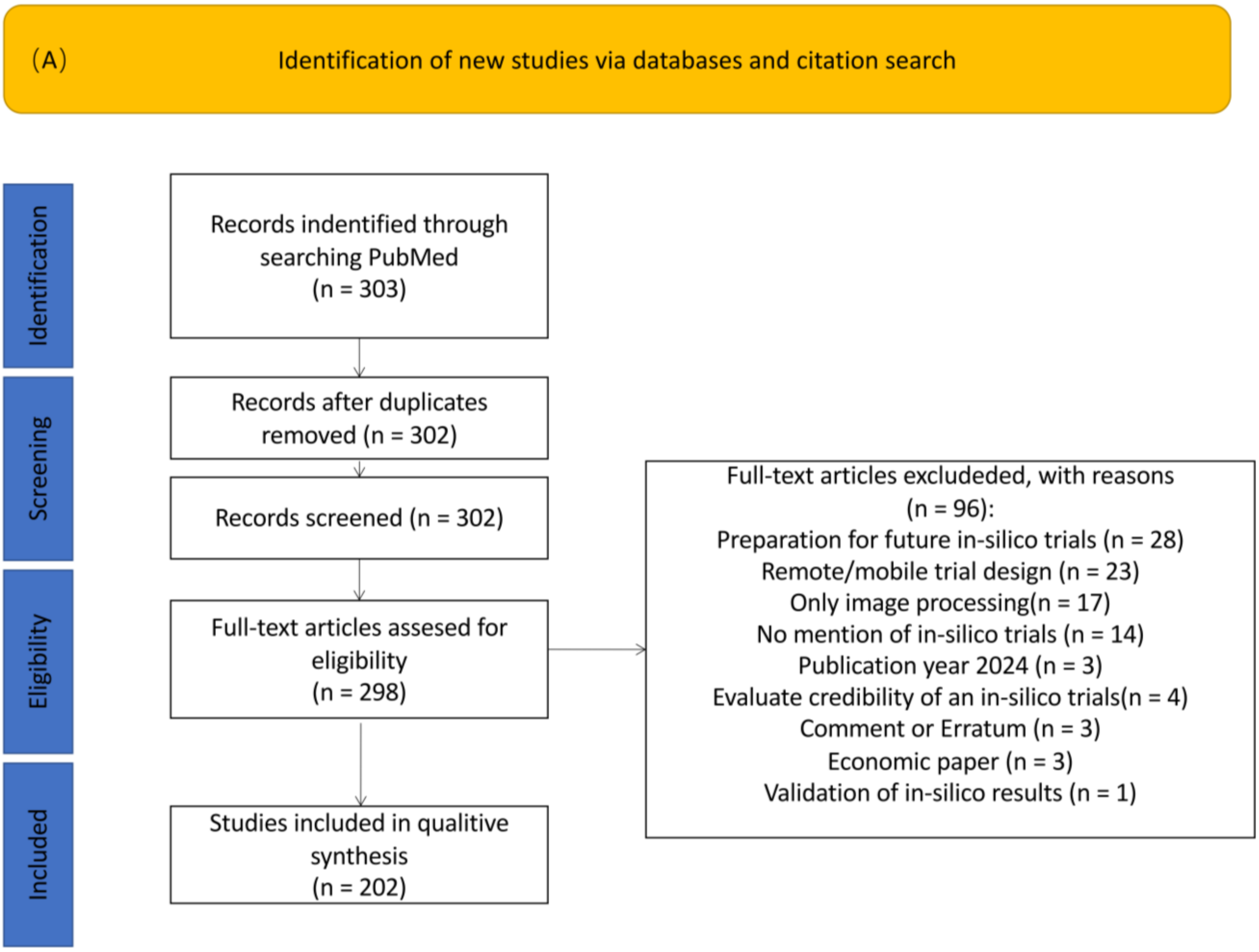

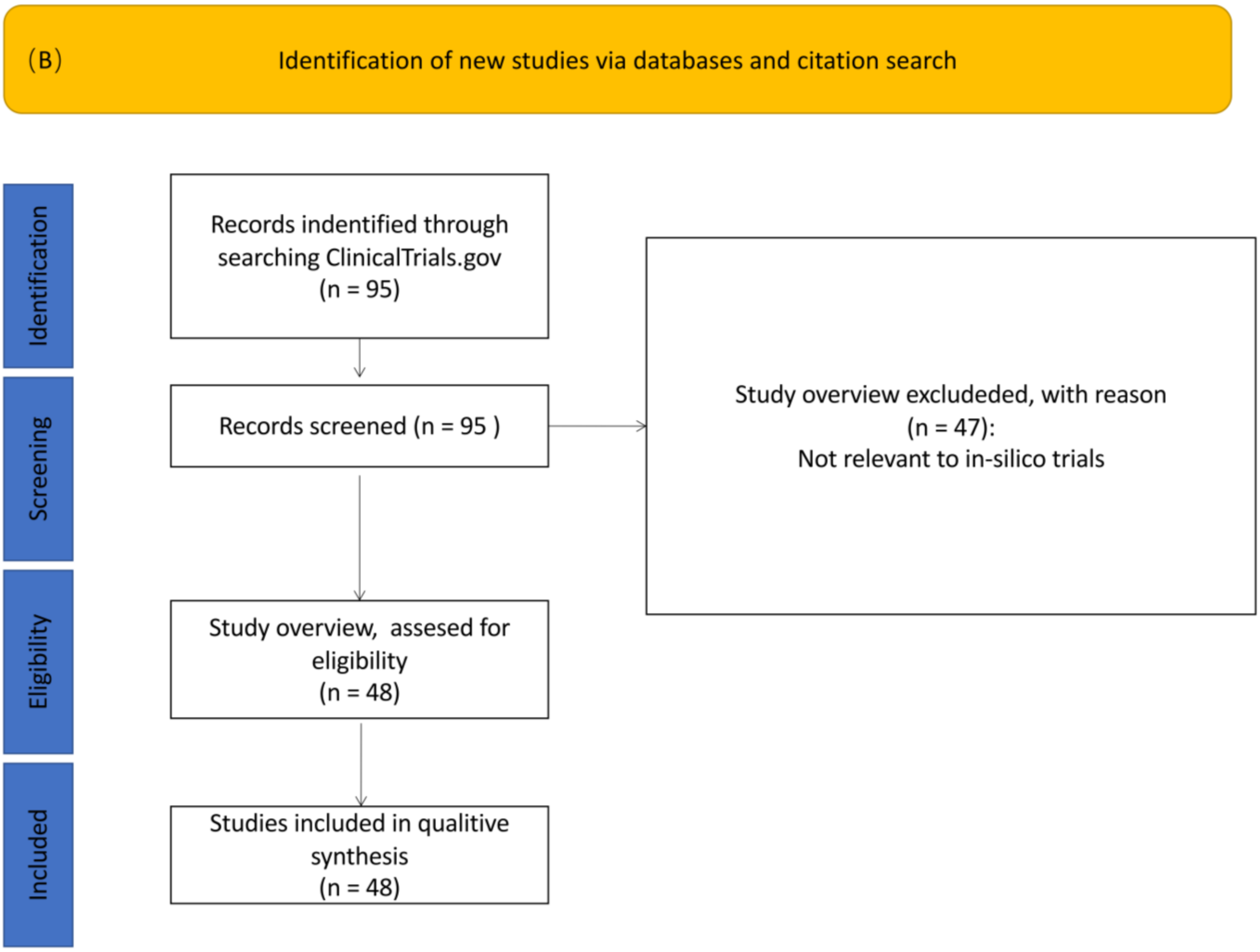

Figure 1. PRISMA flow charts of PubMed (Panel (A)) and ClinicalTrials.gov searches (Panel (B)).

## Screening and classification results for articles and clinical trials

In the following two subsections, we summarize the numbers of articles and clinical trials that covered the topic of ISCTs along with the classification of their aims.

Screening for relevance and classification of purpose

Figure 2 illustrates the focus regarding ISCTs for the two searches in PubMed and ClinicalTrials.gov. In the PubMed data set, the majority of articles is categorized as "*IS application articles"* (54%), indicating a significant emphasis on practical application of IS methods. *"IS methods articles"* account for 31.2%, reflecting interest in methodological advancements, while "*IS review articles*" represent the smallest share at 14.9%. In registered trials, the category *"IS applications"* makes up 52.1%. Additional categories (*"Preparation for future ISCTs*", 35.4%, and *"Validation of ISCT results",* 12.5%) emphasize ongoing development and validation efforts. Overall, the results from the PubMed and ClinicalTrials.gov search both demonstrated a stronger focus on IS applications. In the following, we point out some typical examples in the different categories of registered trials. An example of a trial categorized as *"Validation of ISCT results"* is NCT02651181 [20], which tested a real, physical system called the Hybrid Logic Closed Loop (HLCL) designed to help people with type-1 diabetes better control their blood sugar levels. Initially, the system was evaluated through computer simulations, demonstrating its ability to reduce both low (hypoglycemia) and high (hyperglycemia) blood sugar episodes and stabilize glucose levels. The HLCL system operates by using a Continuous Glucose Monitor (CGM) to track blood sugar in real-time, sending data to a control algorithm that calculates the required insulin dose and instructs an insulin pump to deliver it, automatically regulating blood sugar through a feedback loop. This trial was categorized as *"Validation of ISCT results"* because its primary goal was to confirm the effectiveness and safety of the system, already validated through IS testing, in a real-world environment. Rather than introducing a new intervention, the study aimed to ensure that the simulation results translated effectively into practical, supervised clinical settings.

The *"Preparation for the future ISCTs"* category includes trial NCT04663828 [21], which focuses on finding better ways to treat tinnitus, a condition that causes ringing or buzzing in the ears. The trial looks at different treatment options, comparing single treatments with combinations and targeting both the ear and brain to find the most effective approach. To support future ISCTs, the study collects data from clinical trials and combines it with information about patients' genetics, proteins, and hearing tests. Using this data, researchers build computer models that can simulate how different treatments might work in the future. These models are part of a decision support system designed to help doctors choose the best treatment plans, ultimately aiming to improve outcomes for people with tinnitus and to guide the design of future ISCTs.

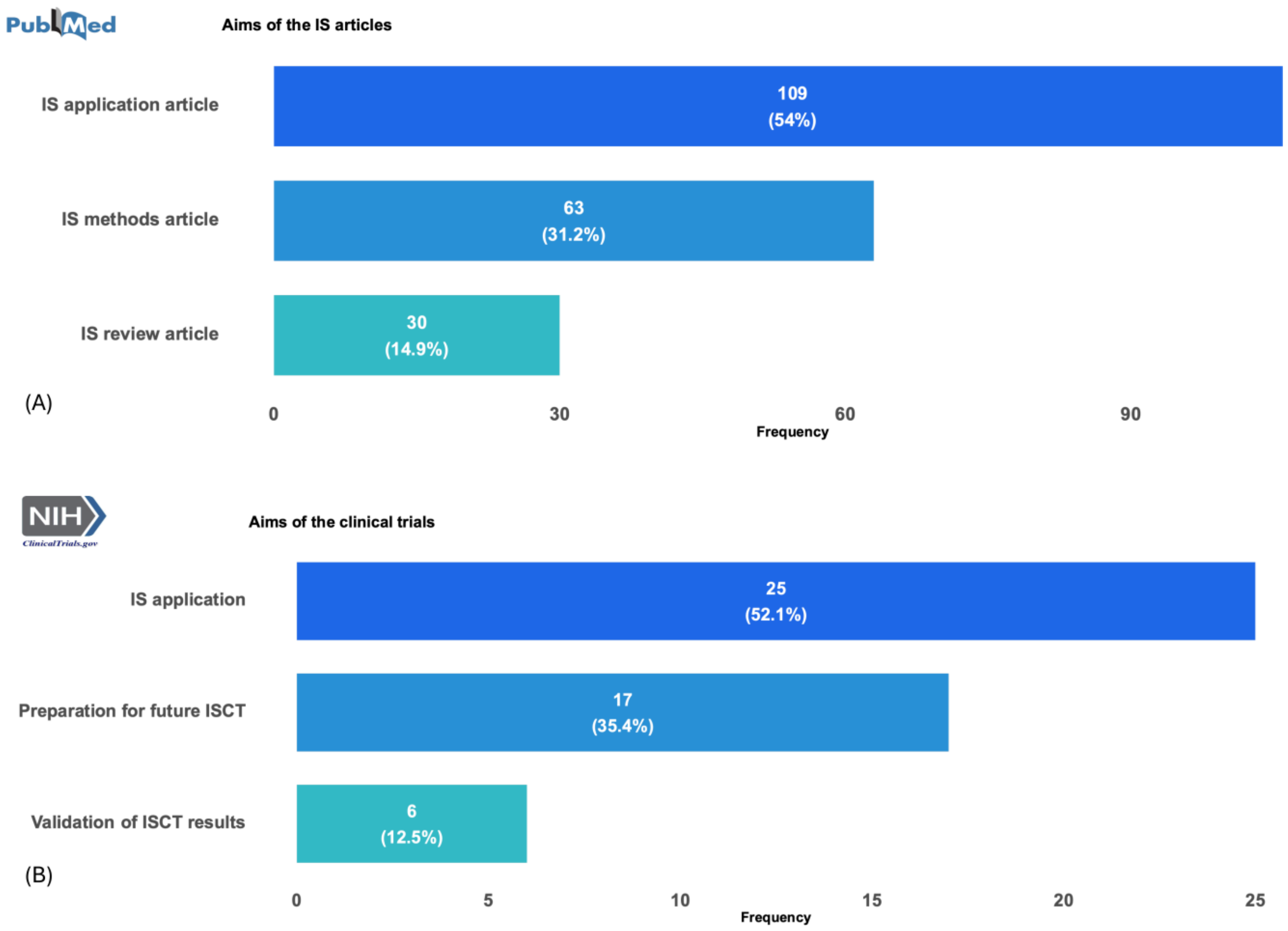

Figure 2: Classification by connection to IS methods for publications (Panel (A)) and clinical trials (Panel (B)).

<u>Relation to drug development</u>

Figure 3 shows the classification of drug development. Of the 202 articles from the PubMed search, 76 (37.6%) focused on specified compounds, 37 (18.3%) discussed drug development without specifying a compound, and 89 (44%) were unrelated to drug development completely. The 48 trials from the ClinicalTrials.gov search featured a similar distribution: 19 (39.6%) focused on specified compounds, 7 (14.6%) did not specify a compound, and 22 (45.8%) were unrelated.

Some articles and trials did not specify particular drugs or treatments and focused on broader objectives, such as developing computational models or general methodologies, which can later be applied to specific therapies.

While a considerable portion of articles and trials is not directly tied to drug development, a notable number still focuses on this area. This is because, beyond drug development, many ISCTs also explore fields like imaging and diagnostics. Nevertheless, drug development clearly remains a major focus, emphasizing its current strong relevance to and strong potential for future ISCT applications.

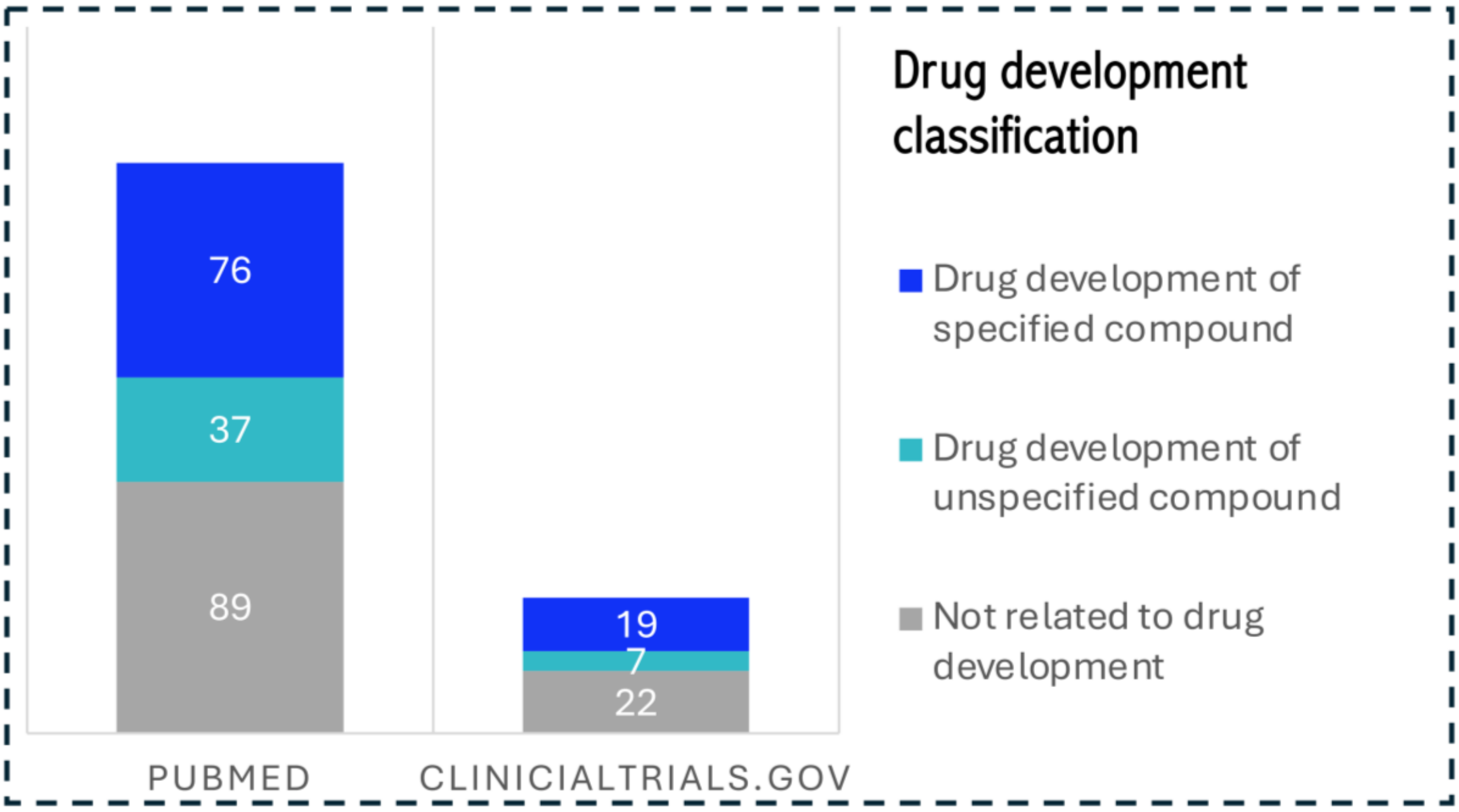

Figure 3. Drug development classification for the searches in PubMed and ClincialTrials.gov.

**Time Trends**

Figure 4A and 4B show the proportions of relevant journal articles and trials in our review by year. Over the past years, an absolute and relative increase in articles and trials referring to IS approaches can be observed. There was a slight decline and fluctuation in 2020-2022, likely due to the impact of the COVID pandemic, however, a substantial increase again is evident in 2023, and trends are broadly similar for publications and trials. It is apparent that an increasing number of clinical trials involve the use of IS approaches.

**Geographic Distribution**

Figures 4C and 4D show the geographic distribution of first authors' affiliations and the countries in which the clinical trials are conducted. The majority of publications originate from Europe and the United States, followed by authors from Canada, Brazil, Australia, Japan and China (see Figure 4C). In contrast, the registered clinical trials (Figure 4D) exhibit a more limited geographic distribution, primarily concentrated in the United States and Europe, indicating that while ISCTs and approaches are globally recognized in research, their practical implementation in clinical trials remains more regionally focused.

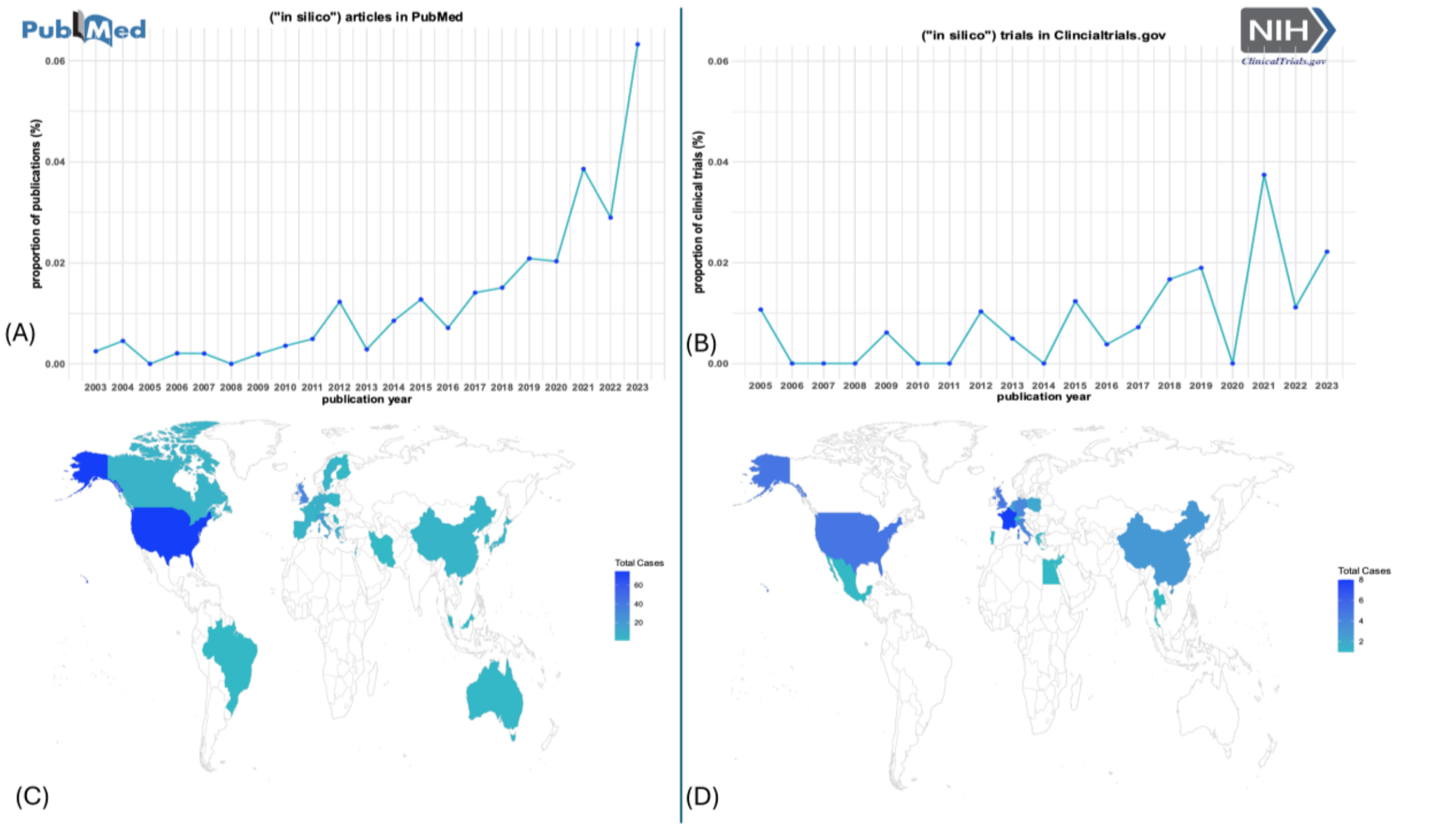

Figure 4. Meta-data on articles included in the systematic review. Panels (A) and (B) show the increase in numbers of articles and trials relating to IS approaches over the years. Panel (C) displays the geographic distribution of the journal articles' first authors' affiliations. Panel (D) displays the countries in which the clinical trials are conducted.

### Disease classification

Based on the ICD-11 disease classification system, Figure 5 compares the application areas of ISCTs in journal articles and registered clinical trials. The analysis highlights notable differences between academic research and practical applications in clinical trials, particularly in the range of diseases covered and the focus of research efforts. While journal articles tend to explore a broader spectrum of diseases, clinical trials currently exhibit more limited coverage, indicating a gap between research and clinical implementation.

The data reveal a strong prevalence of neoplasms in both data sets, with journal articles showing a higher percentage of specific drug development—33 (16.3%) for specified drug compounds and 10 (5.0%) for unspecified compounds—compared to clinical trials, which report only 4 (8.3%) for specified and 2 (4.2%) for unspecified drug compounds. Despite significant academic interest in neoplasm drug development, the number of registered clinical trials remains low (4, 8.3%), possibly due to limited sample sizes in these studies. Additionally, there is a notable research focus on factors influencing health status or contact with health services, covered by 45 (22.3%) journal articles, though no corresponding clinical trials have been registered. Discrepancies also appear in other areas: mental health research has 8 (4%) articles on specific drug developments but no registered trials; infectious or parasitic diseases are covered in 10 articles (5%) versus just 1 trial (2.1%); and respiratory diseases appear in 7 articles (3.5%) compared to 2 trials (4.2%). The "general diseases" category (14 articles, 6.9%) includes cases where the exact disease was unspecified. In these instances, IS approaches may focus on broader biological systems, such as the immune system, rather than specific diseases. Naturally, such a category does not appear in clinical trials, as they require a clearly defined disease focus.

Both searches identified ICD-11 extension codes [22], which provide additional diagnostic context (e.g., Histopathology) and cover the innocence of the adverse events, highlighting the broad application of IS approaches across diverse conditions. Published articles encompass most disease types based on ICD-11 classifications. In contrast, clinical trials cover fewer disease categories, indicating the potential for future expansion of IS approaches into a broader spectrum of diseases.

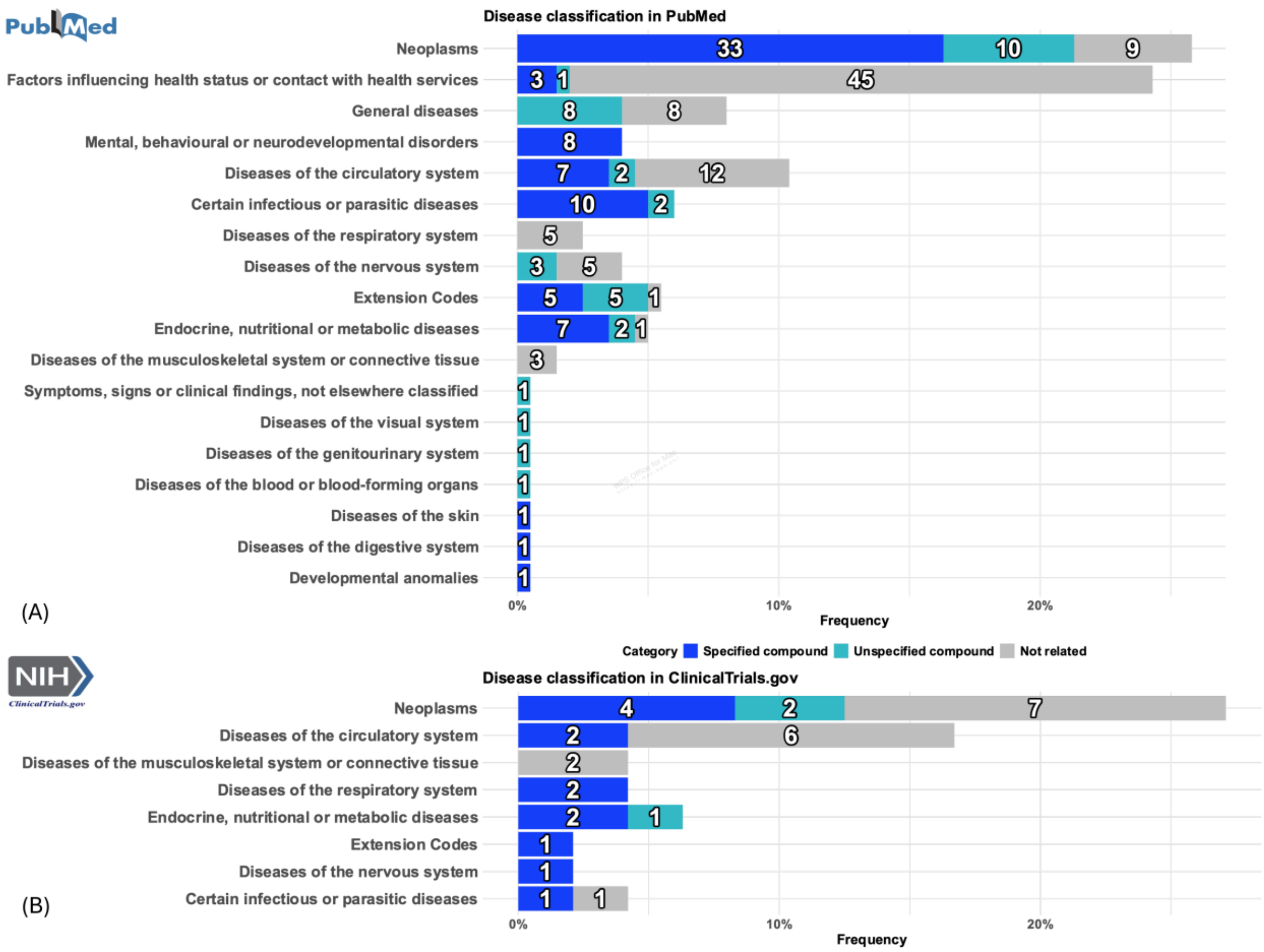

Figure 5. Overview of disease classifications, as well as rare and pediatric diseases presented in articles and trials. Panels (A) and (B) show the disease classification according to ICD-11 for journal articles and clinical trials. We differentiate between “not related to drug development”, “related to drug development but not to a specific drug compound” and “related to a specific drug compound”. The x-axis is displayed as a percentage, with a denominator of 202 and 48, representing the total number of included articles and trials from PubMed and ClinicalTrials.gov.

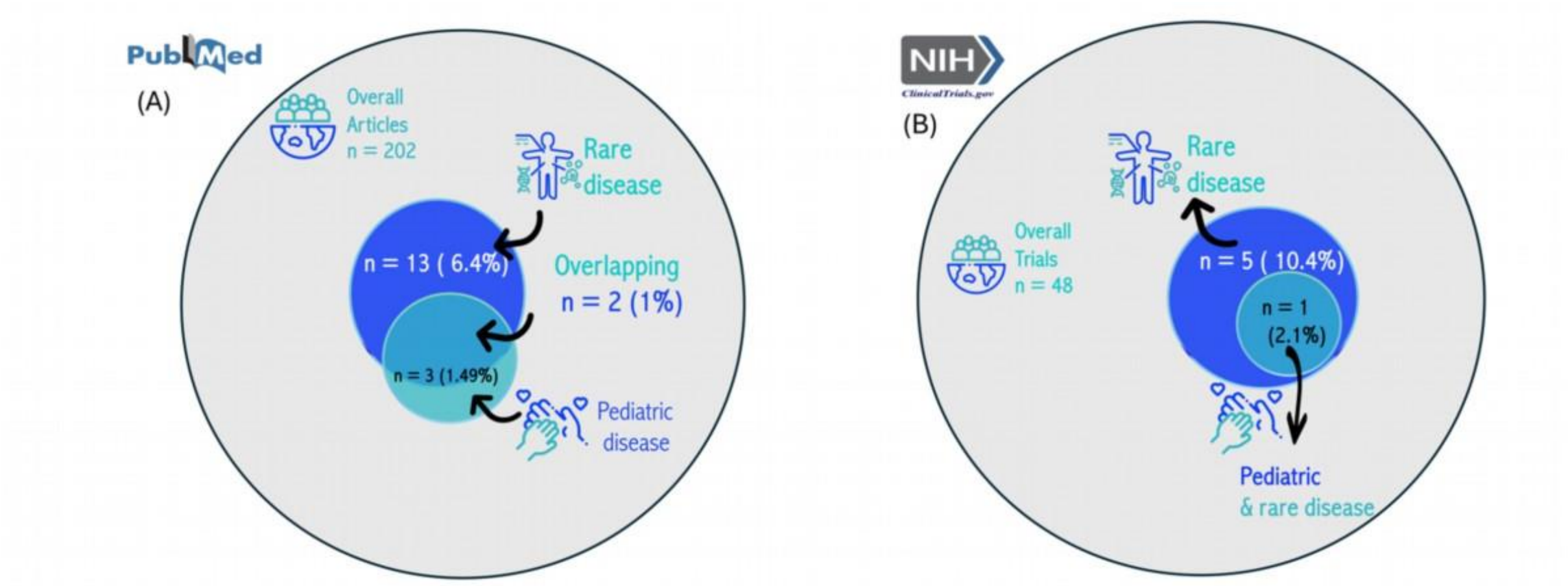

Figure 6. Panels (A) and (B) showthe number of rare and pediatric diseases out of all the pulications and trials.

Examples of IS approaches in the context of rare or pediatric diseases

ISCTs are applicable across various disease contexts, and their ability to simulate patient outcomes rather than relying solely on direct observation can significantly advance research in clinical areas that often face challenges like small sample sizes or ethical concerns, such as in rare or pediatric diseases. To provide insight into how different IS methods are applied in these contexts, we summarize two exemplary articles and one clinical trial below, all of which specifically correspond to the overlapping area of rare diseases and pediatric diseases shown in Figure 6 (A) and Figure 6 (B).

- Example 1: In silico clinical trials for pediatric orphan diseases

The article by Carlier in 2018 [23] presents the implementation and analysis of an investigative ISCT aimed at testing the effects of bone morphogenetic protein (BMP) treatment on congenital pseudarthrosis of the tibia (CPT), a rare pediatric disease. Research on CPT is hindered by small sample sizes and ethical concerns, resulting in limited data and treatment options. The ISCT was carried out using 200 virtual subjects, each one characterized by a set of 8 variables, e.g. Fibroblastic proliferation, Fibroblastic differentiation and Cartilage formation.For each virtual subject, outcomes with and without BMP treatment were simulated using a previously developed and validated mechanistic model of bone regeneration in mice. This model, based on partial differential equations, captures complex non-linear interactions between variables. The results demonstrated a significant reduction in CPT severity following BMP treatment, though considerable variability among subjects was observed. To further analyze these differences, the authors developed a machine learning model that identified three distinct subpopulations and pinpointed key predictive factors. This machine learning model, a random forest model, was trained on 75% of the data with 10-fold cross-validation and the accuracy was assessed on the remaining 25% of the data. Although the ISCT provided valuable insights into CPT treatment, the authors noted limitations, including time-intensive simulations and reliance on a mouse-based model due to missing clinical data on human bone regeneration.

- Example 2: Overcoming chemotherapy resistance in low-grade gliomas: A computational approach

Delobel (2023) [24] reported on the development and application of a mathematical model describing how low-grade gliomas (LGG), a rare form of cancer, respond to temozolomide (TMZ) chemotherapy. The primary goal was to identify new TMZ administration protocols that could improve survival, reduce toxicity, and delay the onset of TMZ resistance. To achieve this, LGG patients exhibiting TMZ resistance were recruited, and an ISCT was conducted using virtual twins and a virtual cohort. The mathematical model, based on a set of ordinary differential equations, was fitted to and validated with longitudinal volumetric MRI data. After estimating individual tumor behavior parameters, virtual twins were generated to explore various TMZ administration schemes. This approach was extended to a virtual cohort by inferring the joint distribution of parameters from real data and sampling virtual subjects. The mathematical model was then applied to the virtual cohort, testing different TMZ regimens and analyzing the resulting survival curves.

This article comprehensively illustrates the role of ISCTs in clinical research and the integration of diverse data types. Initially, a clinical trial was conducted, and the collected data, combined with mechanistic knowledge, was incorporated into a mathematical model. This model was then utilized within an ISCT to simulate new data, which could be used to explore treatment outcomes and potentially guide future clinical research.

- Example 3: A new posaconazole dosing regimen for paediatric patients with cystic fibrosis and aspergillus infection (cASPerCF)

The cASPerCF study (NCT04966234) [25] is a registered multi-center clinical trial that began in 2021, aiming to provide new insights into Aspergillus infections in children and adolescents with Cystic Fibrosis (CF). One of the trial's objectives was to propose and investigate an IS modeled dose of posaconazole, based on a previously published dosing algorithm. To further explore PK/PD and assess the efficacy of the proposed dosing regimen, a randomized open-label study involving 135 children and adolescents with CF and Aspergillus infection, aged between 8 and 17 years, was planned. Unfortunately, the current status of the study is unknown, and no results or methodological details, e.g. how the dosing regimen was modeled, are currently available.

**ISCTs in drug development**

In the following, we consider publications related to drug development, which make up 76 articles (37.6%) among our relevant search results (see Figure 3). We begin by examining the types of MIDD models used in these articles, then analyze the data sources used for model calibration and validation, and finally assess the reproducibility of the data and implementation.

Popular models in drug development

As pointed out in the terminology section, the model definitions in this context do not allow for a distinct model classification; therefore, we relied on the self-labelled categories provided by the original authors. In most articles, the authors reported the developed or applied models as belonging to the categories of PBPK (18 articles, 23.7%), QSP (18 articles, 23.7%) or PK/PD (12 articles, 15.8%); see Figure 7. These models are all referenced in the list of existing and proposed guidelines of the ICH MIDD roadmap, Table 1 in [5]. Moreover, 8 articles (10.5%) reported the integration of artificial intelligence (AI) methods with these modeling frameworks.

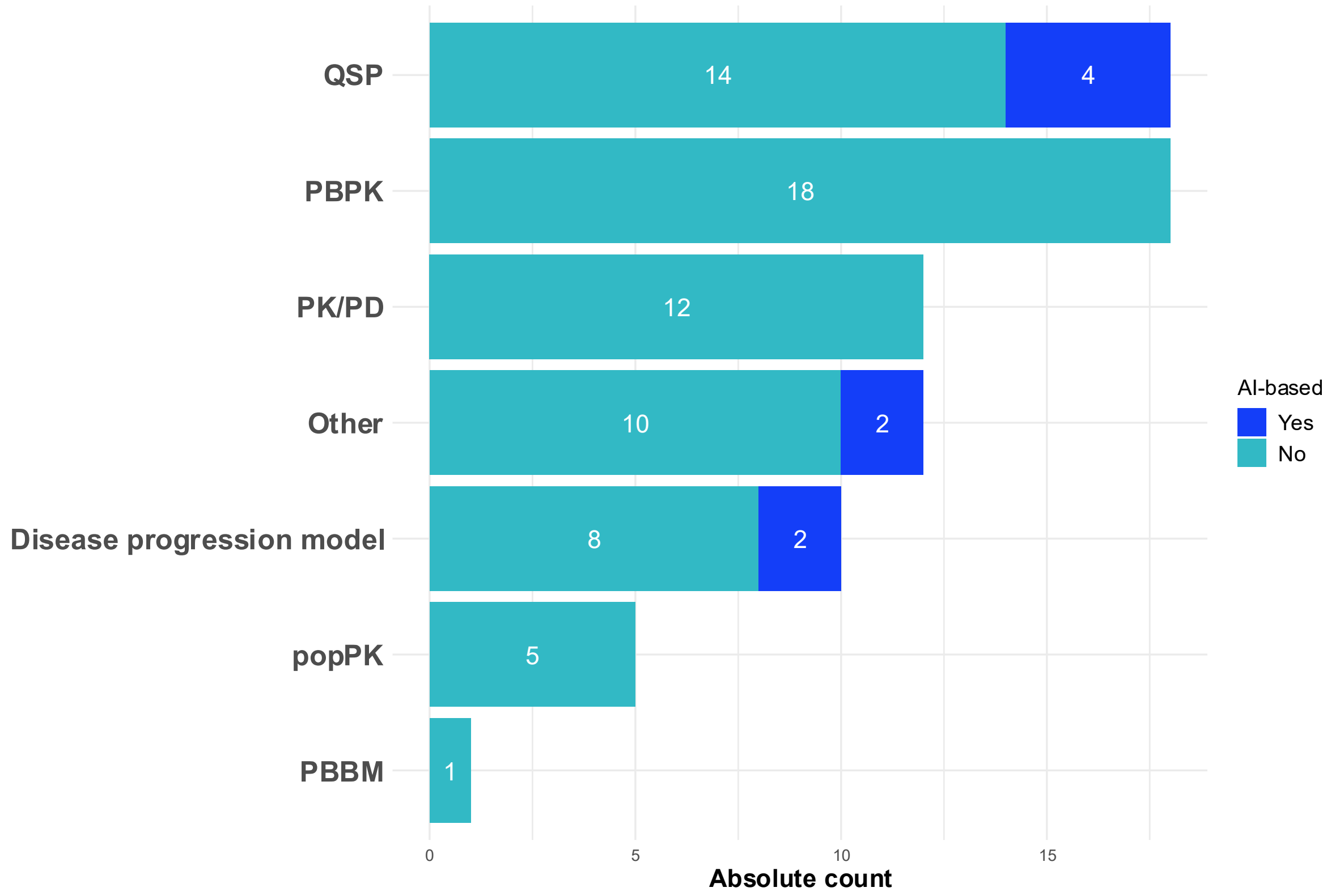

Figure 7. Classification of model informed drug development approaches used among those publications reporting in the context of drug development, shown as a stacked bar chart including AI integration

Data sources of model validation and calibration in PubMed

Next, we take a closer look at the 76 drug-development-related articles to examine the data sources used for model validation and calibration. We present our result in the form of a complex upset plot [26], shown in Figure 8, which describes the intersections between data sources used for model validation and calibration. The most commonly used data source is clinical data, with 42 articles using it for calibration and 46 for validation, followed by pre-clinical data, used in 26 articles for calibration and 9 for validation. The most frequently applied approach relies solely on clinical data for both calibration and validation, followed by the use of only pre-clinical data. The third most common method combines both clinical and pre-clinical data. Additionally, some approaches integrate clinical data with either mechanistic knowledge or registry population data sets. This indicates that clinical data remains crucial when developing IS models.

An interesting observation regarding model calibration [28, 29] is that even if we cannot directly estimate model parameters but are aware of the model structure, we can "calibrate" the parameters to align model predictions with observed data. This is particularly valuable in models that incorporate mechanistic knowledge alongside clinical and preclinical data, as it ensures the model's outputs accurately reflect real physiological processes, improving scientific validity and credibility. Moreover, even when not all parameters of a mechanistic model can be estimated directly, researchers can still calibrate the model so that its predictions match observed data, provided the model structure is well understood. This reveals the critical role of mechanistic knowledge in refining and validating IS models.

It is also worth noting that simulation data were used in only a few cases, exclusively for model validation. This demonstrates its limited yet specific role in the validation process. Unlike real clinical data, simulation data are generated based on assumptions and existing models, which introduces inherent biases, making it unsuitable for calibration, where accurate parameter estimation relies on real-world data. Our review further supports this, as none of the models used simulation data for calibration purposes (Figure 8), underlining its restricted role in model validation while acknowledging its limitations.

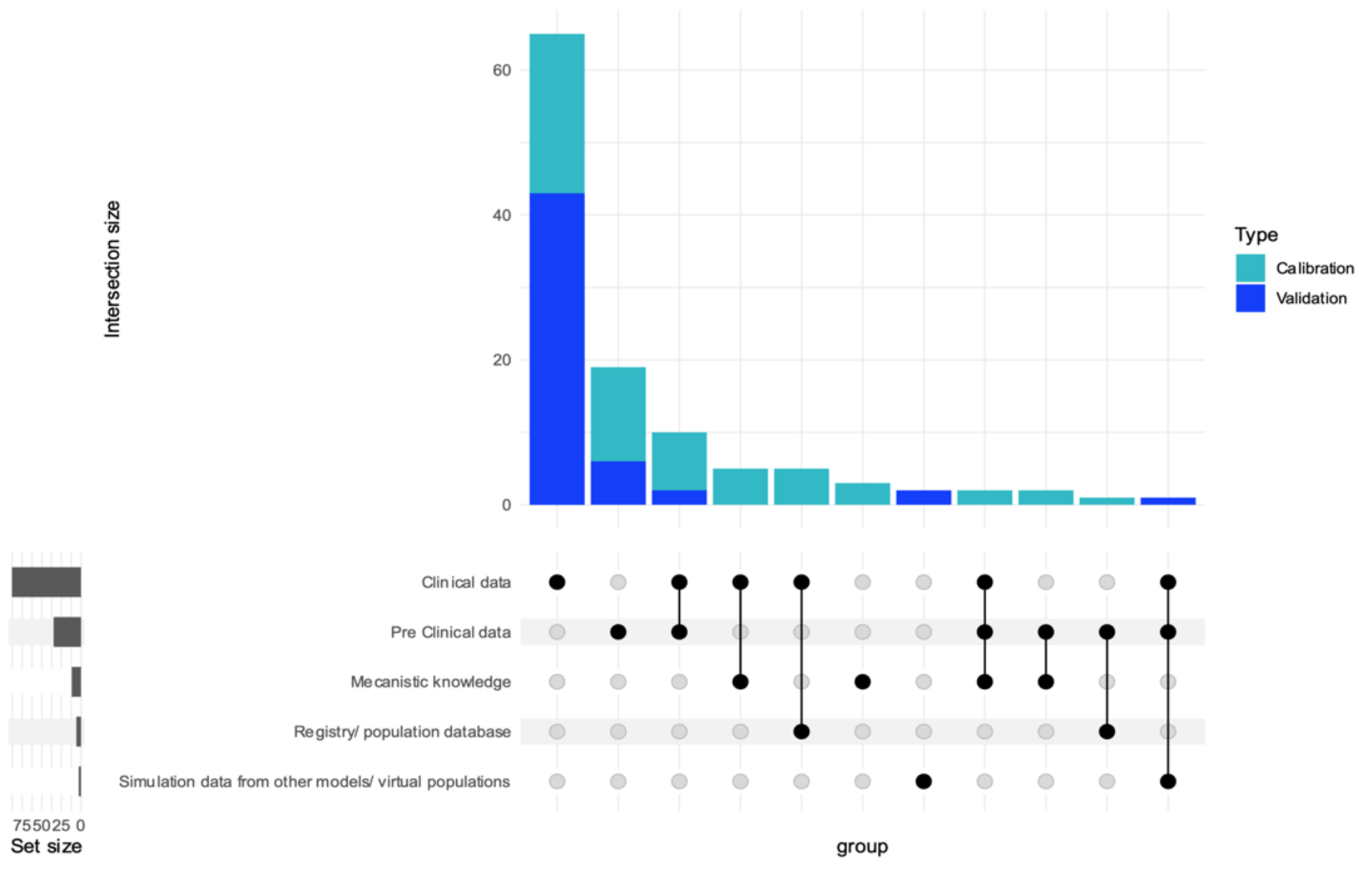

Figure 8. Input type used for the drug  development related IS articles. Displayed are the absolute number a respective input type was used (lower left bars) and the most common combinations of input types (upper bar plot). Model type of calibration or validation is stratified according to the legend in the left.

## Reproducibility in drug development

Figure 9 illustrates the extent to which data and implementations are available in articles related to drug development. Among the journal articles reviewed, 54% developed or applied models that are either open-source (23.7%) or commercially available (30.3%). In 27.6% of the articles, at least the theoretical framework is fully described, allowing for numerical assessment of the model, e.g., evaluating time complexity. However, in 18.4% of the articles, neither the code nor the model equations were shared.

Regarding data availability, the focus was on the resulting data, i.e., data generated from the models rather than the data used for parameter fitting. Only 19.7% of the articles made the generated data publicly available, and in 6.5% of cases, the authors stated that data was accessible upon request, leaving 73.6% of the articles without accessible data or at least unknown status of data availability.

These findings suggest that full reproducibility is not possible in most of the reviewed cases. Nonetheless, the presence of both commercial and open-source solutions shows their established role in model-informed drug development.

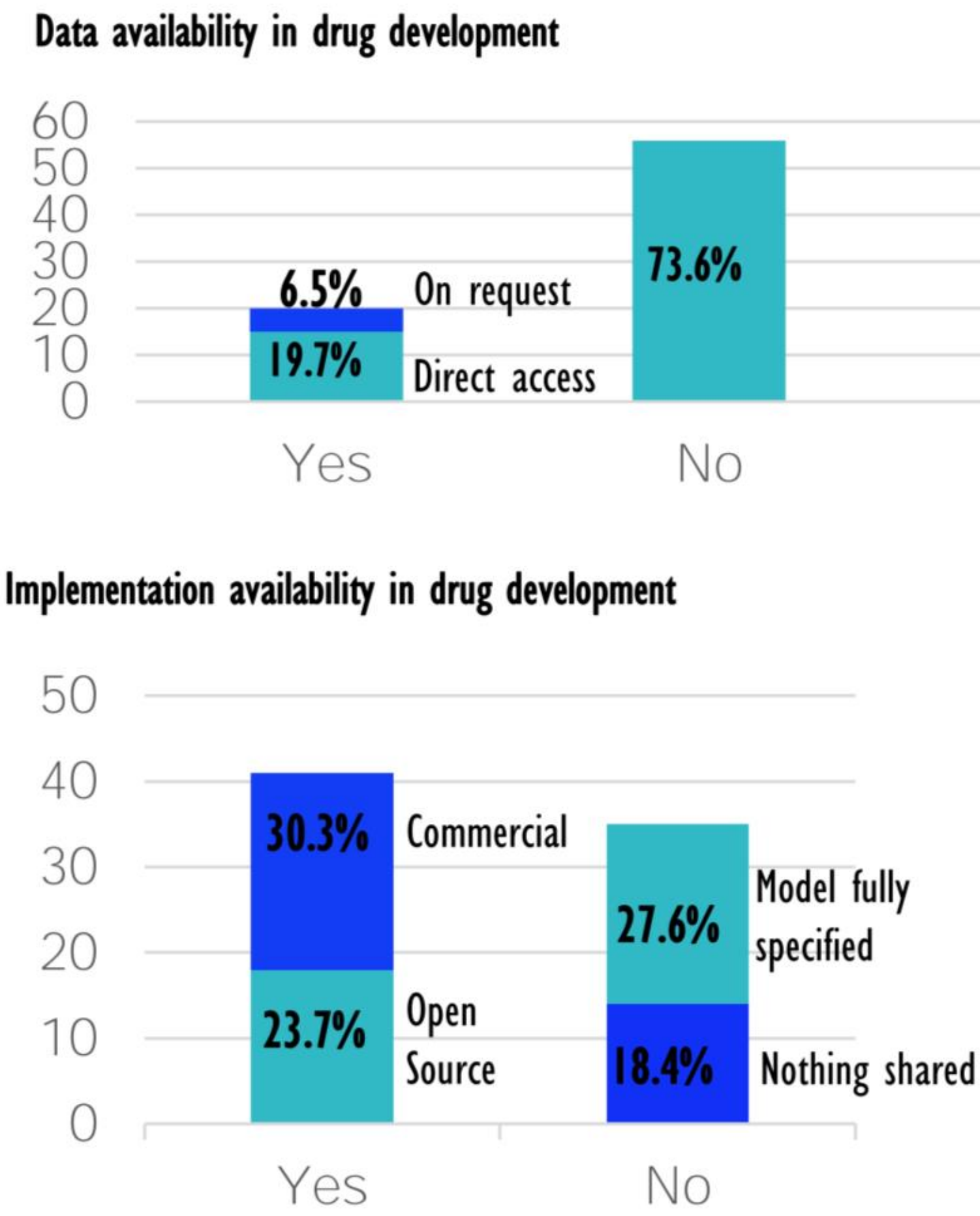

Figure 9. Illustration of the availability of data and implementation in publications related to drug development based on the PubMed search.

### Overlap between journal articles and registered trials

In the journal articles, we searched for NCT identifiers to identify references to ClinicalTrials.gov registry entries, and the corresponding trials along with their characteristics are summarized in Supplementary material S1. Notably, there was no overlap between these trials and those identified through our ClinicalTrials.gov search. Conversely, in the ClinicalTrials.gov registry entries of the identified trials, we examined the cited literature references (PMID identifiers), with the publication details shown in Supplementary material S2. Similarly, no overlap was found between these publications and those identified through our literature review.

Most articles corresponding to the PMIDs cited by the trials focused on the disease or earlier trials involving the same drug rather than the IS aspects, which explains why they were not discovered in the PubMed search.

## DISCUSSION

With data as well as computational resources becoming increasingly accessible, efforts to support or replace expensive physical experiments by simulations have been gaining attention [1, 2, 10]; in this present investigation, we looked into the role of ISCTs in clinical medicine. Use of the term *"in silico"* meanwhile is common in a range of areas, also including, for example, in trials where computational methods are used for diagnostic image processing, or where patients are able to participate remotely. Here we focused on the case where patient data are generated computationally, an important application case that is explicitly considered in several regulatory documents [9, 10]. Our database searches confirmed that reference to IS methods is increasing globally, so far mostly in published articles, but also in registered clinical trials. In our searches, we also did not find overlap between the search results from both domains, which may not be too surprising given the still small body of literature, while it may also hint on a gap between methodological research and practical applications. On the other hand, it is not quite obvious to what extent classical trial registration would be applicable to ISCTs.

Most of the IS models that have been suggested are data-driven, utilizing empirical clinical or preclinical data for calibration or validation, and as such mostly serve to complement rather than to fully replace clinical data. Many IS applications are in the context of cancer research, in imaging or diseases of the circulatory system. Based on our framing (see Introduction), ISCTs appear particularly valuable in the context of rare diseases and pediatric diseases, where traditional clinical trials are often hindered by recruitment difficulties, feasibility issues, and ethical considerations [23–25, 27]. The alignment between the inherent advantages of IS approaches and the structural challenges of rare and pediatric disease trials suggests that these populations could particularly benefit from further methodological development and implementation of ISCTs. The absence of many rare and pediatric diseases in our systematic review likely reflects the limited data available to support model development, calibration, and validation. As our findings show, clinical data are widely used in these processes, highlighting their critical role in enabling reliable in-silico modeling. Although only a small number of ISCT applications in rare and pediatric diseases were identified in our review, this relative scarcity may itself highlight a gap and therefore a strong opportunity for future research. In addition to rare and pediatric diseases, our results also suggest that neoplasms as well as endocrine, nutritional and metabolic diseases, especially diabetes are currently among the most prominent application areas, both in terms of publication volume and model availability. Of the 18 open-source models identified in our sample, 16 were used in neoplasm-related studies. Among 23 studies that employed commercial models, 4 targeted diabetes-related outcomes, indicating that industry-grade tools are beginning to support disease areas with large public health impact.

Regarding transparency and reproducibility of the implemented approaches, the IS models are usually fully specified, while in a majority of cases the implementation is not freely available (or only via commercial software). Despite the absence of privacy concerns with artificially generated outcomes, data sets are only available in a minority of cases. In order to foster applications of IS methods, more transparency and additional efforts to reproduce and validate results might be in order.

Despite these insights, several limitations of our review should be acknowledged. First, the use of restricted search terms and databases may have affected the comprehensiveness of our results. In particular, by focusing on the term "*in silico clinical trial*," we may have excluded relevant studies that use broader terms such as "*in silico*" alone. For instance, the published work on Paxlovid's development [32], which employed IS approaches for compound identification, was not retrieved in our search despite its relevance, as it does not explicitly include the full term "*in silico clinical trial*." While this focused terminology ensures conceptual clarity and consistency, it may also have limited the comprehensiveness of our review. Also, while we distinguished and investigated the various types of models discussed, we relied on the authors' terminologies here and did not re-assess or classify IS models by ourselves.

The present investigation provides an overview of the current state of ISCTs and allows for important insights into their use in clinical medicine, example cases, and promising future research areas. Although the field remains in its early stages, with practical implementations lagging behind methodological advances, it shows strong potential for future growth and standardization. Key aspects to achieve broader acceptance of ISCTs are the availability of validated models [2] as well as their implementation and the resulting data to facilitates reproducibility and transparency.

## CONFLICT OF INTEREST

All authors declared no competing interests for this work.

## AUTHOR CONTRIBUTIONS

All the other authors have participated in the systematic reviews of the IS papers and clinical trial registry; they have edited and reviewed the manuscript.

## FUNDING

This work was partially funded by the DZHK (German Center for Cardiovascular Research), and partially by the European Union's Horizon Europe Framework program under grant agreement 101136365 (INVENTS) , co-funded by the Swiss State Secretariat for Education, Research and Innovation (SERI) and co-funded by the UKRI Innovative UK under their Horizon Europe Guarantee scheme.